\documentclass{article}

\usepackage{amsmath}
\usepackage{graphicx}
\usepackage{dcolumn}
\usepackage{bm}

\usepackage{arxiv}

\usepackage[utf8]{inputenc} 
\usepackage[T1]{fontenc}    
\usepackage{hyperref}       
\usepackage{url}            
\usepackage{booktabs}       
\usepackage{amsfonts}       
\usepackage{nicefrac}       
\usepackage{microtype}      
\usepackage{lipsum}		
\usepackage{graphicx}
\usepackage{natbib}
\usepackage{doi}

\title{Towards Critical Branching Mechanism in Recurrent Neural Networks}

\author{ \href{https://orcid.org/0009-0000-5859-8701}{\includegraphics[scale=0.06]{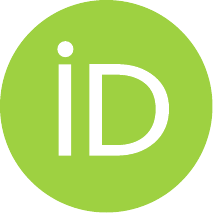}\hspace{1mm}Feixiang Ren} \\
	Department of Physics,\\
	National University of Singapore\\
	Singapore \\
	\texttt{e1353365@u.nus.edu} \\
	\And{Ling Feng} \\
	Institute of High Performance Computing\\
    Agency for Science, Technology and Research (A*STAR)\\
	Singapore\\
	\texttt{0@criticality.ai} \\
}

\begin{document}
\maketitle

\begin{abstract}
	Criticality has been proposed as a key organizing principle in biological neural systems, yet its origin and relevance in artificial neural networks remain unclear. We analyze hidden-state dynamics in trained long short-term memory (LSTM) networks and show that small networks near their optimal training epochs (steps) exhibit scale-free avalanche statistics and branching parameters close to unity, indicative of near-critical dynamics, while larger models remain subcritical. To explain the coexistence of subcritical branching with robust $1/f^{\beta}$ noise, we introduce a mixture branching process framework that links heterogeneous branching dynamics to long-range temporal correlations. These results identify critical-like behavior in LSTMs as an emergent, capacity-dependent dynamical regime. 
\end{abstract}


\section{Introduction}
Scientists have long drawn inspiration from biological neural systems in the design of Artificial Neural Networks (ANNs), with an emphasis on mimicking their structural and functional characteristics. By investigating the dynamical behavior of biological brains, researchers seek to identify principles that can be transferred to ANN design~\cite{hopfield1982neural, rumelhart1986learning}. One key finding in this area is that both biological brains and ANNs tend to operate near the edge of chaos, a critical state that maximizes information-processing capacity~\cite{mora2011biological, bertschinger2004edge, boedecker2012information, feng2019optimal, wei2025multiple}. Motivated by the discovery of neuronal avalanches in biological systems~\cite{beggs2003neuronal}, scale-free avalanche-like activity propagating across layers has also been observed in deep feedforward neural networks such as ResNet \cite{ghavasieh2025toward, he2016identity}. Furthermore, accumulating evidence suggests that both biological neural systems and Long Short-Term Memory (LSTM) networks~\cite{hochreiter1997long} display $1/f^{\beta}$ noise in their activity, a statistical hallmark often associated with self-organized criticality (SOC)~\cite{linkenkaer2005breakdown, montez2009altered, pathania2021exploring, allegrini2009spontaneous, chong2024self, bak1988self, per1987self}.  

In the context of neuroscience, the SOC hypothesis has been extensively investigated, particularly in in-vitro cortical networks, where neural activity organizes into scale-free cascades known as neuronal avalanches  \cite{beggs2003neuronal, plenz2021self, yu2014scale, petermann2009spontaneous}. These avalanches are characterized by power-law distributions of event sizes and durations, indicating the absence of characteristic scales. From a theoretical perspective, such dynamics can be modeled using a mean-field sandpile model, which is mathematically equivalent to a critical branching process \cite{christensen1993sandpile, zapperi1995self}. Within this framework, the branching parameter serves as a control variable that quantifies the balance between activity propagation and decay, with criticality emerging when the system is poised between quiescence and runaway excitation~\cite{harris1963theory}. 

While the SOC paradigm and avalanche dynamics in biological neural systems are well established, the relevance to ANNs remains far less understood. In particular, although recent studies have reported the presence of $1/f^{\beta}$ noise and long-range temporal correlations in LSTM networks \cite{chong2024self}, a unifying theoretical framework explaining the emergence of such scale-free behavior in trained networks is still lacking. Unlike biological systems, ANNs are deterministic, parameterized models trained via gradient-based optimization, raising fundamental questions about how critical-like dynamics arise in the absence of explicit self-organizing mechanisms.

Here, we propose a framework to extend the concept of neuronal avalanches from neuroscience to the realm of ANNs, with a particular focus on LSTM networks. By defining avalanche-like events in LSTM hidden-state dynamics and analyzing their statistical properties, we establish a direct connection between LSTM avalanches and the branching process. Furthermore, we demonstrate that the emergence of $1/f^{\beta}$ noise in LSTMs is closely linked to the underlying branching structure of network dynamics, providing a possible mechanistic interpretation of scale-free temporal correlations in LSTMs.

The remainder of this paper is organized as follows. In Sec.~\ref {LSTM networks}, we introduce the LSTM networks, dataset, and training protocols in our analysis. In Sec.~\ref{Avalanches}, we define avalanche events in LSTM dynamics and demonstrate the emergence of the scale-free avalanche size distribution at specific stages of training and for particular network configurations. In Sec.~\ref{Branching process}, we adopt a branching-process framework to account for the observed avalanche behaviors and apply an established method to estimate the branching parameter, which serves as a quantitative indicator of the dynamical state of the network. However, the branching-process analysis in Sec.~\ref{Branching process} primarily captures short-range temporal correlations and does not account for the long-range $1/f^{\beta}$ noise observed in LSTM activity. To address this limitation, in Sec.~\ref{Long-range dynamics} we investigate the long-range temporal correlations in LSTMs and introduce an extended branching mechanism capable of reproducing the observed $1/f^{\beta}$ behavior. This framework yields an ensemble-averaged branching parameter whose evolution mirrors that of the conventional branching parameter across training epochs and network configurations. Throughout the paper, an epoch denotes a discrete step in the training process corresponding to one full pass over the dataset.

\section{LSTM networks}
\label{LSTM networks}
\begin{figure}[t]
    \centering
    \includegraphics[width=0.5\linewidth]{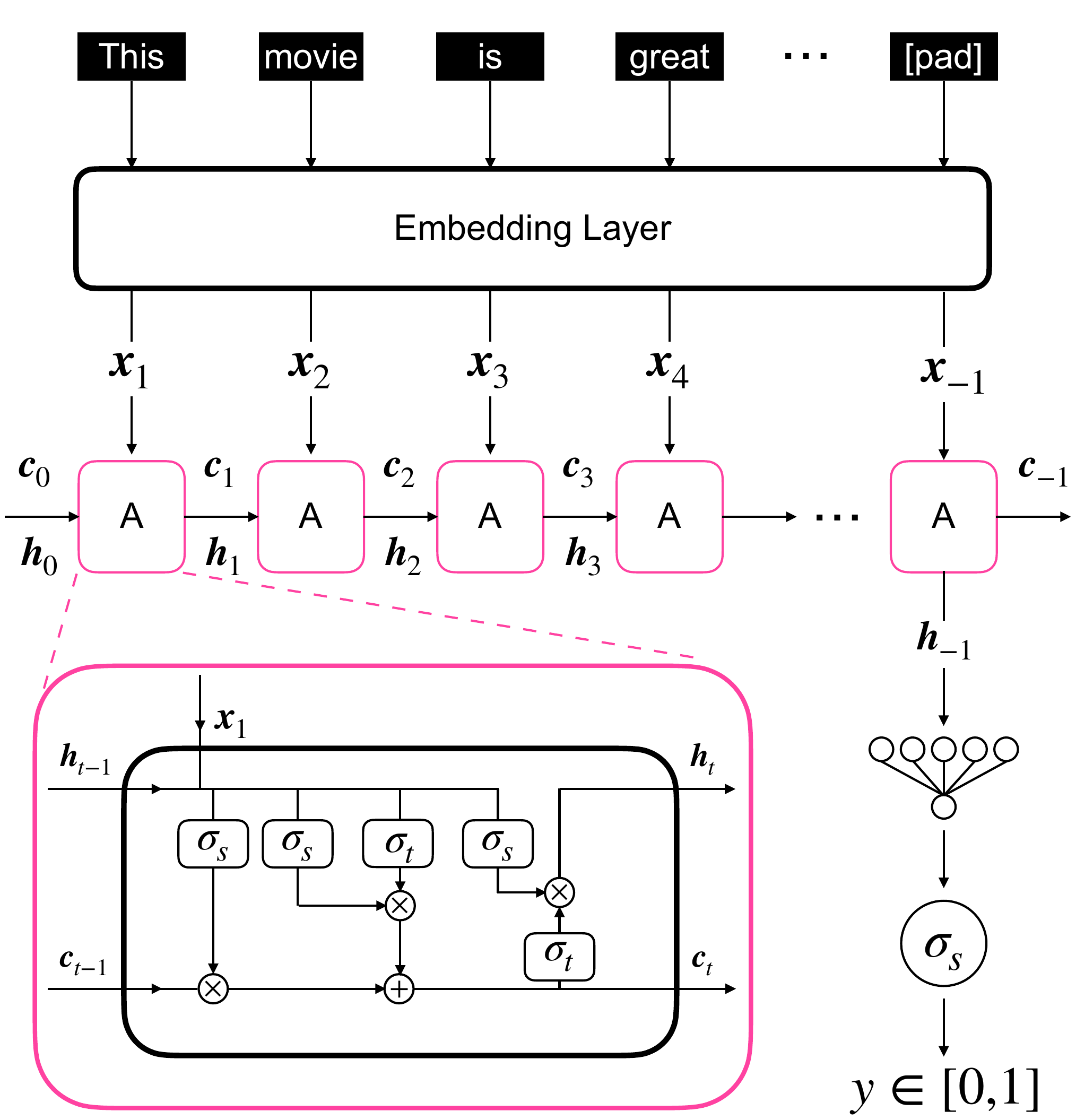}
    \caption{Schematic of the LSTM architecture used for binary sentiment classification. Block $A$ corresponds to the LSTM layer, with the internal gating and state-update operations shown in the pink inset.}
    \label{fig1}
\end{figure}
The LSTM architecture employed in this work consists of three components: an embedding layer, a single LSTM layer, and a linear classifier implemented as a dense layer followed by a sigmoid function, as shown in Fig.~\ref{fig1}. The embedding layer maps each discrete input token onto a continuous representation in an eight-dimensional (vector) latent space. The single-layer LSTM can be mathematically described as~\cite{hochreiter1997long}:
\begin{equation}
    \label{equation1}
    \begin{aligned}
        \boldsymbol{i}_t &= \sigma_s\left(\boldsymbol{W}_i\boldsymbol{x}_t + \boldsymbol{U}_i\boldsymbol{h}_{t-1} + \boldsymbol{b}_i\right), \\
        \boldsymbol{f}_t &= \sigma_s\left(\boldsymbol{W}_f\boldsymbol{x}_t + \boldsymbol{U}_f\boldsymbol{h}_{t-1} + \boldsymbol{b}_f\right), \\
        \boldsymbol{o}_t &= \sigma_s\left(\boldsymbol{W}_o\boldsymbol{x}_t + \boldsymbol{U}_o\boldsymbol{h}_{t-1} + \boldsymbol{b}_o\right), \\
        \tilde{\boldsymbol{c}_t} &= \sigma_t\left(\boldsymbol{W}_c\boldsymbol{x}_t + \boldsymbol{U}_c\boldsymbol{h}_{t-1} + \boldsymbol{b}_c\right), \\
        \boldsymbol{c}_t &= \boldsymbol{f}_t \odot \boldsymbol{c}_{t-1} + \boldsymbol{i}_t \odot \tilde{\boldsymbol{c}_t}, \\
        \boldsymbol{c}_{\mathrm{out}t} &= \sigma_t\left(\boldsymbol{c}_t\right), \\
        \boldsymbol{h}_t &= \boldsymbol{o}_t \odot \boldsymbol{c}_{\mathrm{out}t}, \\
    \end{aligned}
\end{equation}
where $\boldsymbol{x}_t$ denotes the eight-dimensional embedding vector corresponding to the input token at timestep $t$, $\odot$ indicates element-wise multiplication, $\sigma_s(x)=\frac{1}{1+e^{-x}}$ and $\sigma_t(x)=tanh(x)$ are the sigmoid and hyperbolic tangent functions, respectively, and $\left\{\boldsymbol{W}, \boldsymbol{U}, \boldsymbol{b}\right\}$ represent the trainable parameters of the LSTM layer. At each timestep $t$, the embedded input $\boldsymbol{x}_t$ is fed into the LSTM layer together with the previous hidden state $\boldsymbol{h}_{t-1}$ and cell state $\boldsymbol{c}_{t-1}$, which store the memory of the network, resulting in an updated hidden state $\boldsymbol{h}_t$ and cell state $\boldsymbol{c}_t$. After the entire review is processed by the LSTM layer, the final hidden state $\boldsymbol{h}_{-1}$ is passed to the classifier, which outputs a value in the interval $[0,1]$ for binary classification.

\begin{figure}[t]
    \centering
    \includegraphics[width=0.6\linewidth]{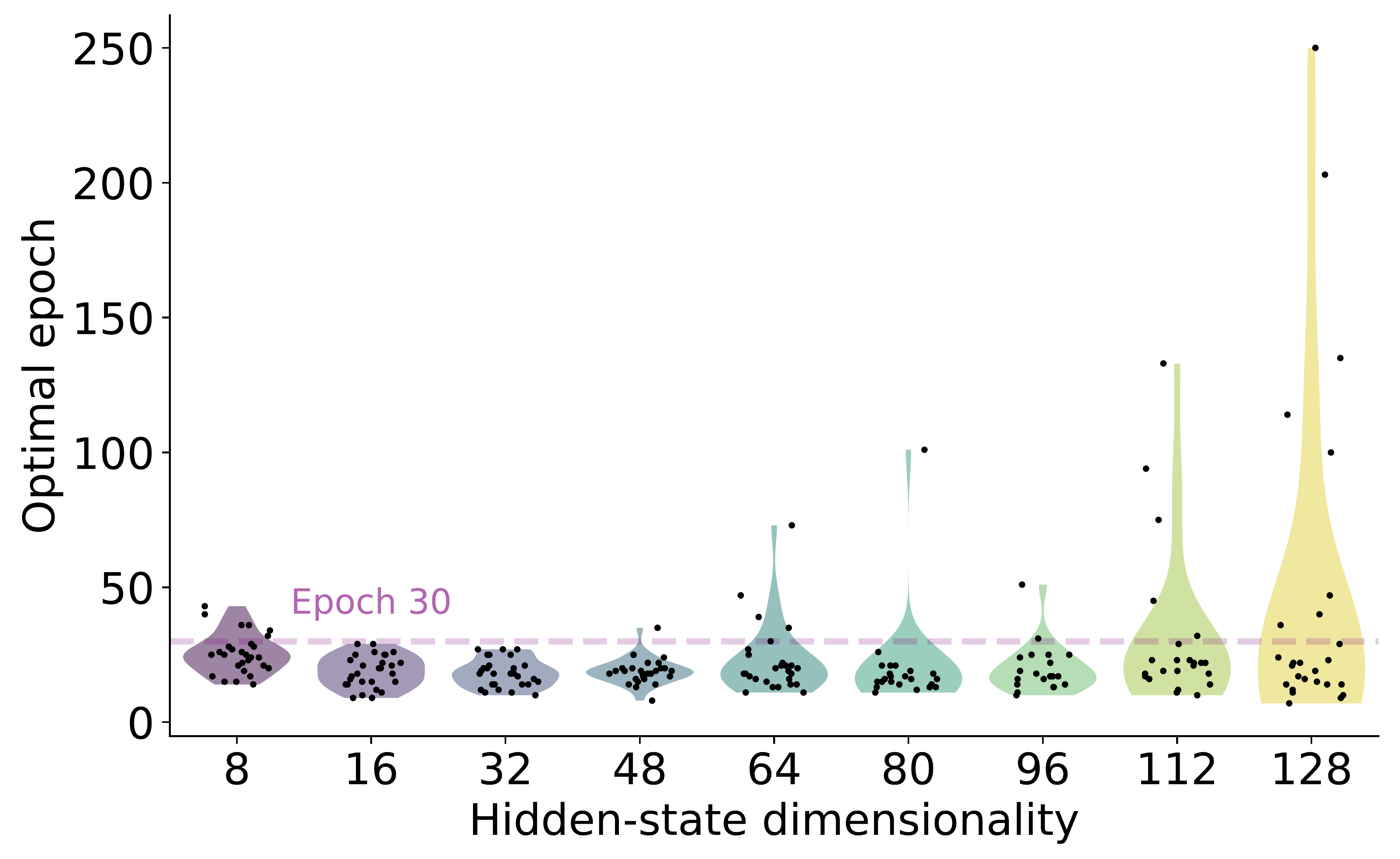}
    \caption{Distribution of optimal epochs (i.e., the epoch with the lowest test loss during training) as a function of hidden-state dimensionality for LSTM networks. Each violin plot summarizes the distribution of optimal epochs across multiple network realizations with different random seeds but identical architectures. Black dots denote individual realizations. The dashed horizontal line marks epoch $30$, which corresponds to the epoch at which the majority of networks achieve their best performance across hidden-state dimensions.}
    \label{fig11}
\end{figure}

We trained multiple LSTM networks sharing the same architecture described above, while varying the dimensionality of the hidden state $\boldsymbol{h}_t$ across the set $\{8, 16, 32, 48, 64, 80, 96, 112, 128\}$. The networks were trained on a dataset of $50,000$ IMDb movie reviews~\cite{maas2011learning}, labeled as positive opinion (1) or negative opinion (0). To account for the inherent stochasticity in neural network training and to ensure the statistical robustness of the results presented in this work, each configuration was trained using $28$ different random seeds; a small number of trained networks were excluded for reasons discussed in Appendix~\ref{Appendix D}. Across all realizations, most networks saturated before epoch $30$ at a test accuracy of approximately $87\%$, whereas larger networks exhibited delayed optimization for a small subset of random seeds, as illustrated in Fig.~\ref{fig11}. All observables measured below were computed during inference using the network checkpoints at the specified epoch. For a given hidden-state dimensionality and training epoch, the same quantity was evaluated across all experiments trained with different random seeds, and the results were averaged, with error bars representing the standard deviation unless explicitly stated otherwise. Although reviews in the dataset varied in length, all were truncated or padded to a fixed length of $500$ tokens. The dataset was then split into a training set of $35,000$ reviews and a test set of $15,000$ reviews. All networks were trained using an identical set of hyperparameters, summarized in Table~\ref{hyperparameters table}.

\begin{table}[t]
\centering
\caption{Selected hyperparameters used during LSTM training.}
\label{hyperparameters table}
\begin{tabular}{ccc}
\toprule
\shortstack{\textbf{Batch size}\\128} &
\shortstack{\textbf{Embedding size}\\8} &
\shortstack{\textbf{Dropout}\\0.1} \\
\midrule
\shortstack{\textbf{L2 regularization}\\0.0005} &
\shortstack{\textbf{Learning rate}\\0.005} &
\shortstack{\textbf{Gradient clipping}\\5} \\
\bottomrule
\end{tabular}
\end{table}



\section{Avalanches}
\label{Avalanches}
In neuroscience, neuronal avalanches are defined as cascades of spatiotemporal bursts of neural activity separated by intervals of quiescence. Experimentally, such avalanches are typically identified using multielectrode arrays placed onto cortical slices, which record local field potentials (LFPs) from multiple electrodes~\cite{beggs2003neuronal, petermann2009spontaneous, priesemann2014spike}. Neuronal populations in the vicinity of a given electrode are considered active within a discrete time bin if a negative spike exceeding a predefined threshold is detected in the corresponding LFP signal during that interval. An avalanche is operationally defined as a sequence of consecutively active time bins, where a bin is considered active if it has at least one active electrode, preceded and followed by inactive bins. The size and duration distribution of the avalanches in cortical slices are observed to be scale-free, a hallmark of critical dynamics in biological neural systems.

\begin{figure*}
    \centering
    \includegraphics[width=\linewidth]{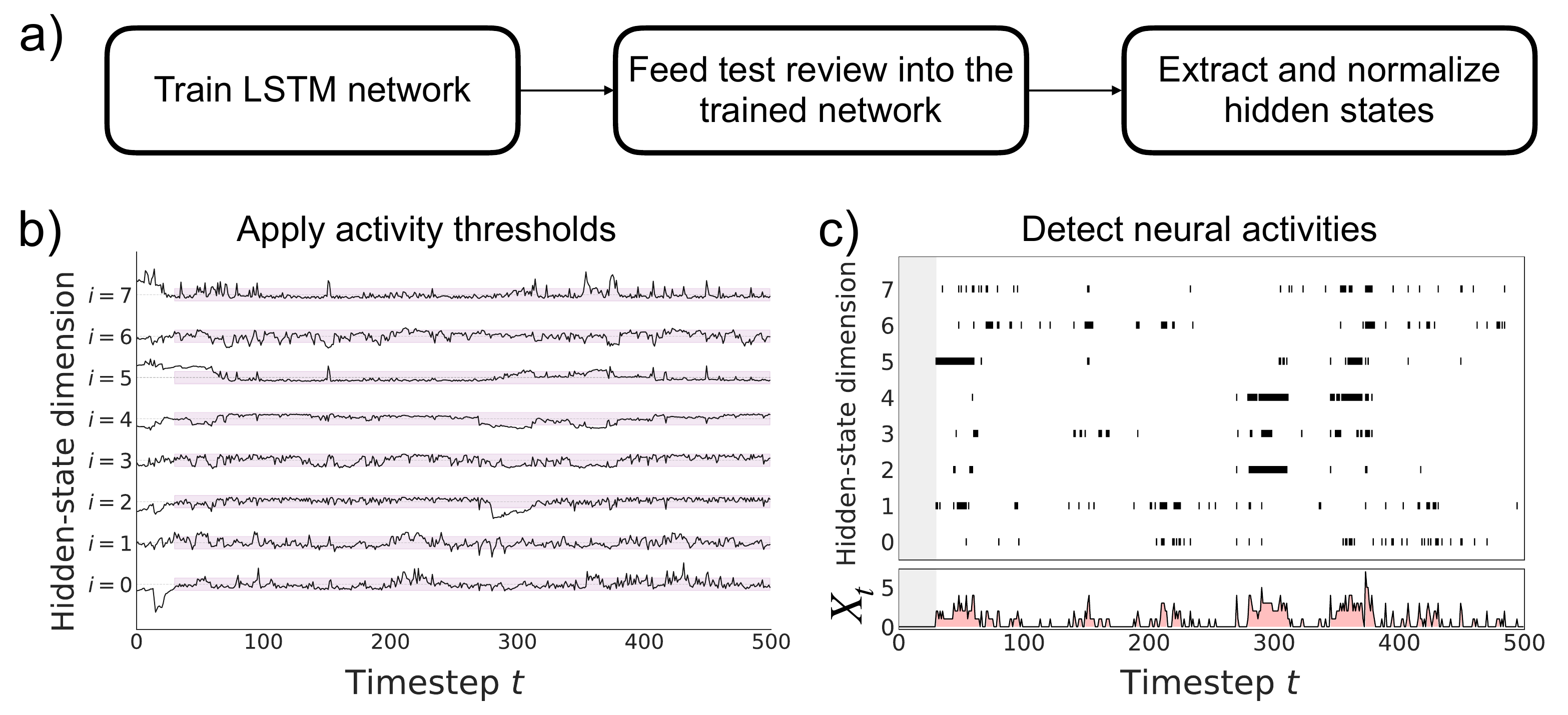}
    \caption{Extraction of neural activities from an LSTM network. (a) Schematic of the analysis pipeline. LSTM processes test reviews to generate hidden states $\boldsymbol{h}_t$, which are recorded and normalized for subsequent analysis. (b) Time series of normalized hidden states for selected dimensions $i$. A uniform activity threshold (shaded regions) is applied independently to each normalized hidden-state dimension to identify significant activation events. For illustration, the hidden-state dimensionality is set to eight; in the experiments, networks with varying hidden dimensions were analyzed. (c) Binary raster plot of detected neural activities across dimensions after thresholding (top), together with the corresponding $X_t$, defined as the number of active neurons at each timestep $t$ (bottom). Consecutive timesteps for which $X_t\neq0$ (red regions) are identified as avalanches.}
    \label{fig2}
\end{figure*}

Motivated by this experimental framework, we modeled each dimension of the hidden states $\boldsymbol{h}_t$ in LSTM as an artificial neuron and applied an analogous avalanche analysis during network inference, as illustrated in Fig.~\ref{fig2}(a). Specifically, reviews from the test set of the IMDb dataset were processed using the trained LSTM networks. For each review, a sequence of 500 tokens was sequentially fed into the LSTM layer, yielding $\boldsymbol{h}_t$ at every timestep. To render the activities of different artificial neurons directly comparable, each dimension of the resulting $\boldsymbol{h}_t$ was independently standardized across time using $z$-score normalization described below. 

\begin{figure*}
    \centering
    \includegraphics[width=0.7\linewidth]{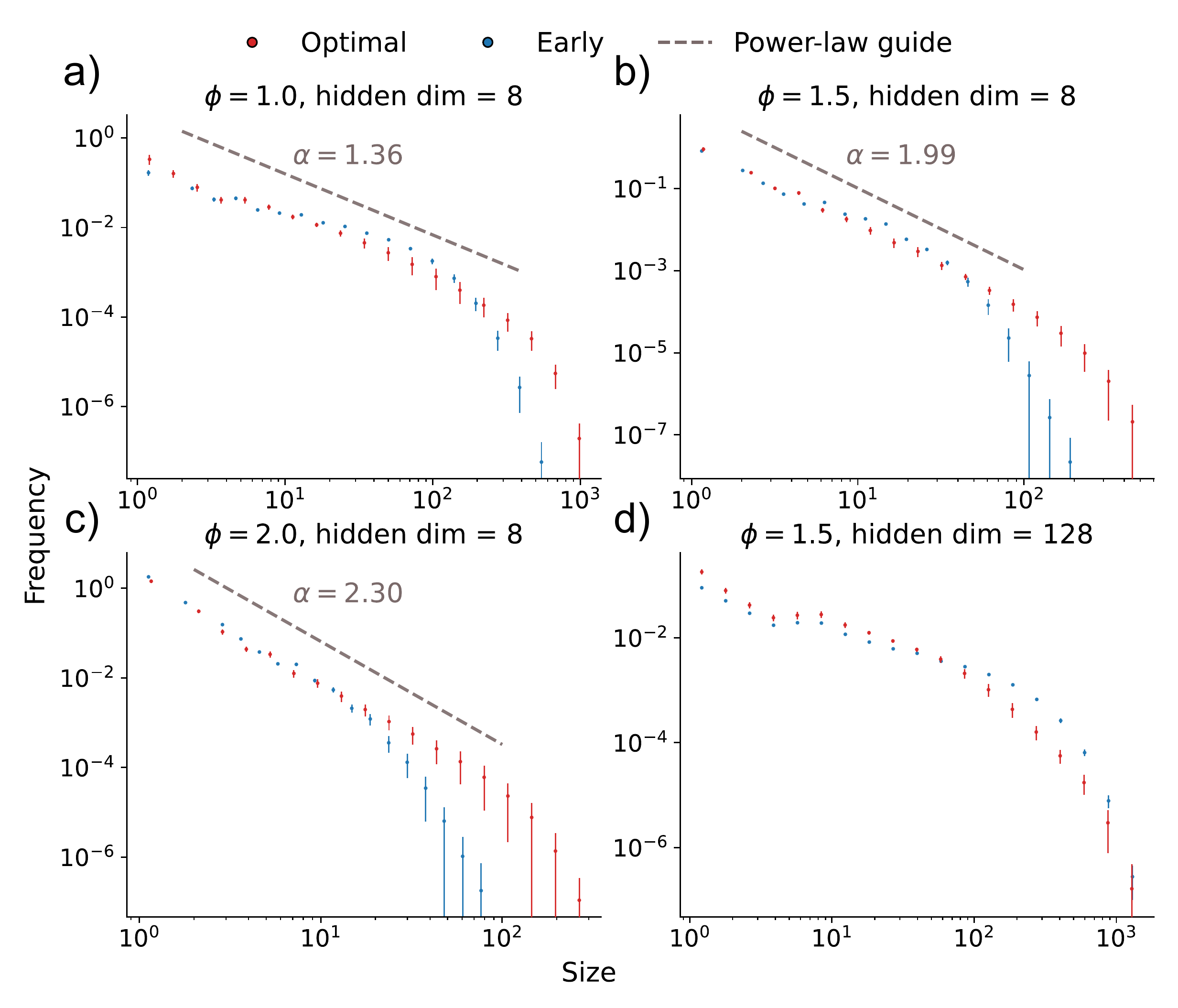}
    \caption{Log–log plots of avalanche size distributions obtained from thresholded hidden-state activities across different operating regimes. Red markers correspond to the optimal training epoch, while blue markers indicate an early training stage. Here, epoch $30$ is selected as a representative optimal epoch across realizations, despite variation in the exact optimal epoch for different networks (Fig.~\ref{fig11}). This definition is used consistently in all subsequent figures unless noted otherwise. Error bars represent statistical variability across networks trained with different random seeds but identical architectures and training epochs. Dashed lines denote power-law guides, fitted to the avalanche size distributions at the optimal epoch, with the corresponding exponent $\alpha$ shown for reference. Panels (a–c) show results for a network with hidden-state dimensionality 8 at increasing activity threshold $\phi$: (a) $\phi=1.0$, (b) $\phi=1.5$, and (c) $\phi=2.0$. Panel (d) shows the avalanche size distribution for a larger network (hidden-state dimension 128) at $\phi=1.5$. Results for additional large hidden-state dimensions, which exhibit similar behavior, are provided in Appendix~\ref{Appendix D}. The observed power-law scaling for small networks remains qualitatively consistent across threshold values $\phi$, supporting the presence of near-critical dynamics in small LSTM networks.}
    \label{fig3}
\end{figure*}

For a given hidden-state dimension $i$, the temporal mean $\mu_i$ and standard deviation of the $i$-th hidden dimension were computed over the interval $t \in [t_0, T]$, where $T=500$. The initial segments $t<t_0$ were discarded as a burn-in period, during which the hidden-state dynamics were strongly influenced by the initialization of the hidden states $\boldsymbol{h}_0$, and cell states $\boldsymbol{c}_0$. The normalized neural activities were defined as the absolute deviation from the temporal mean measured in units of the corresponding standard deviation,
\begin{equation}
    \label{equation1b}
    \begin{aligned}
        \mu_i &= \frac{1}{T-t_0+1}\sum_{t'=t_0}^{T}h_{t',i}, \\
        |h^{\mathrm{norm}}_{t,i}| &= \frac{|h_{t,i} - \mu_i|}{\sqrt{\frac{1}{T-t_0+1}\sum_{t'=t_0}^{T}(h_{t', i}-\mu_i)^2}},
    \end{aligned}
\end{equation}
where $i$ indexes the hidden-state dimension and $T-t_0+1$ denotes the length of the time window used for normalization. This normalization step ensures that differences in activity scale across artificial neurons do not bias the subsequent avalanche detection. As shown in Fig.~\ref{fig2}(b), after normalization, all dimensions of the normalized neural activities $|\boldsymbol{h}^{\mathrm{norm}}_{t}|$ were binarized using a uniform threshold $\theta$, which was manually selected and applied consistently across dimensions and time. An artificial neuron was considered active at a given timestep if the amplitude of its normalized $\boldsymbol{h}_t$ exceeded this threshold, as demonstrated in Fig.~\ref{fig2}(c). This binarization defines a binary state matrix $\boldsymbol{s}_t$ with elements
\begin{equation}
    \label{equation2}
    s_{t,i}=
    \begin{cases}
        1, & \text{if } |h^{\mathrm{norm}}_{t,i}| > \theta, \\
        0 & \text{otherwise}.
    \end{cases}
\end{equation}
A timestep was classified as active if at least one neuron was active at that moment. An avalanche was then identified as a sequence of consecutive active timesteps bounded by silent periods on both sides. The use of $z$-score normalization and threshold-based binarization follows standard practice in studies of neuronal avalanches~\cite{plenz2021self, yu2014scale}, from which our methodology is inspired. This entire pipeline was applied to all 15,000 test reviews, after which we analyzed the avalanche statistics over the entire set.

We analyzed the distribution of avalanche size $P(S)$, where the size $S$ is quantified as the total number of active neurons accumulated over the duration of an avalanche, across networks with different sizes. For the network with the smallest hidden-state dimensionality at its optimal epoch, the avalanche size followed a power-law distribution with an exponential cutoff. The presence of this cutoff reflects finite-size effects, which are expected in any system with finite degrees of freedom, including biological neural systems~\cite{beggs2003neuronal}. As shown in Fig.~\ref{fig3}(a)-(c), within a reasonable range of threshold values $\theta$, the power-law behavior remained qualitatively stable. However, the estimated power-law exponent varied substantially under different thresholding conditions, indicating limited robustness of the avalanche statistics, similar to what has been reported for neuronal avalanches in biological neural systems. In the early epochs, the smallest network typically exhibited a pronounced exponential decay in the avalanche size distribution, suggesting subcritical dynamics~\cite{christensen1993sandpile}. In Fig.~\ref{fig3}(d), for larger networks, the avalanche size generally deviates from the power-law distribution at both early and optimal training epochs. For clarity, only the case with hidden-state dimension $128$ is shown in the figure; corresponding results for other large hidden-state dimensions are provided in Appendix~\ref{Appendix D}. This behavior indicates that large networks remain subcritical during training, with no transition toward criticality observed.

\section{Branching process}
\label{Branching process}
To account for scale-free avalanche statistics, biological neural systems are often modeled as a mean-field sandpile, which can be mathematically described as a branching process governed by an order parameter known as the branching parameter $m$~\cite{beggs2003neuronal, harris1963theory}. The dynamics were described as
\begin{equation} 
    \label{equation3}
    X_{t+1} = \sum_{i=1}^{X_t} Y_{t,i}, \quad \langle Y_{t,i} \rangle=m,
\end{equation}
where $X_t$ denotes the number of toppled sites (or active electrodes) at discrete time $t$, and $Y_{t,i}$ represents the number of offspring events generated by the $i$-th toppled site at that time. The branching parameter $m$ quantifies the expected number of descendants per toppled site. As $m$ approaches one, the sandpile self-organizes into a critical state, where a perturbation can trigger a cascade of events that neither dies out nor explodes. In this critical regime, the avalanche size distribution of this sandpile follows a power law with a critical exponent of $\tau_s=\frac{3}{2}$, independent of the threshold value. Notably, the robustness of the exponent $\tau_s$ in the mean-field theory is considerably stronger than that observed in cortical slice experiments~\cite{beggs2003neuronal}, a discrepancy that is commonly attributed to the idealized connectivity structure of the mean-field model. In contrast, when $m < 1$, cascades eventually dissipate, and the avalanche size distribution will deviate from the power-law behavior. Consequently, measuring the branching parameter $m$ is crucial for understanding the mechanism underlying the avalanche dynamics in LSTM. 

\begin{figure*}[t]
    \centering
    \includegraphics[width=\linewidth]{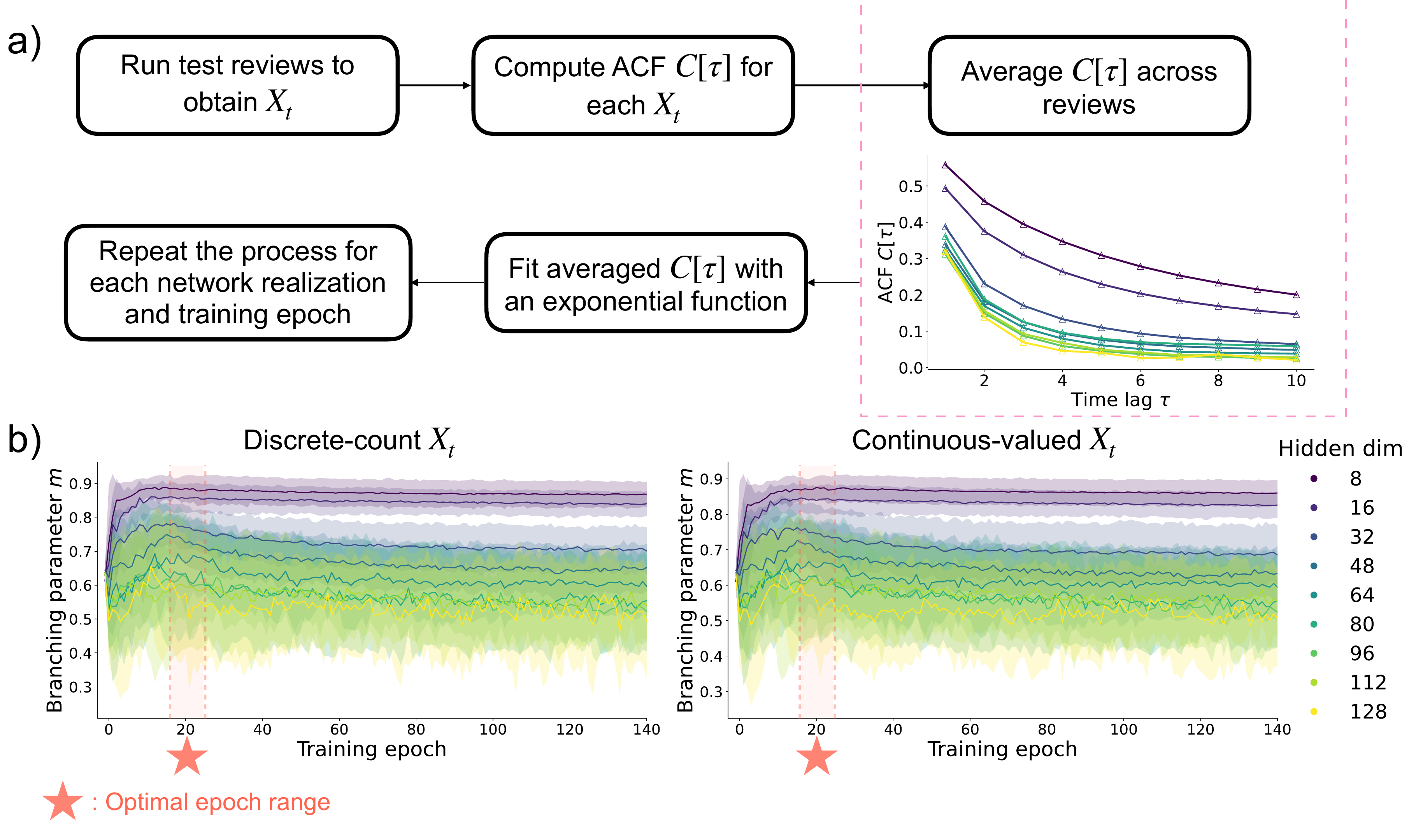}
    \caption{Estimation of the branching parameter from temporal correlations in LSTM activities. (a) Schematic of the analysis pipeline. For each trained network, test reviews are processed to obtain the activity signal $X_t$. The ACF $C[\tau]$ is computed for each signal and averaged across reviews. The averaged ACF is then fitted with an exponential decay to extract an effective branching parameter. This procedure is repeated across networks with different hidden-state dimensions, across independent training runs governed by different random seeds, and across training epochs. (b) Evolution of the estimated branching parameter $m$ over training epochs for different hidden-state dimensions, obtained from discrete-count activity signals $X_t$ (left) and continuous-valued definitions of $X_t$ that eliminate explicit thresholding and binarization of the hidden states $\boldsymbol{h}_t$ (right). Solid lines indicate the mean across networks sharing the same hidden-state dimensionality but trained under different random seeds, while shaded regions denote variability arising from stochasticity in the training process. The red dashed vertical lines and star markers highlight the epoch interval during which the majority of networks achieve their optimal performance. Larger hidden dimensions systematically yield lower effective branching parameters, while training drives small networks toward a near-critical regime.}
    \label{fig4}
\end{figure*}

Unlike the mean-field sandpile model or in-vitro brain systems, which are externally driven only after relaxation, artificial neurons in LSTM are continuously driven by external input tokens. To capture the branching dynamics underlying the LSTM avalanche, we hypothesized a driven variation of the branching process given by the following equation:
\begin{equation}
    \label{equation4}
    X_{t+1} = \sum_{i=1}^{X_t} Y_{t,i} + I_t, \quad \langle Y_{t,i} \rangle=m,
\end{equation}
where $X_t=\sum_is_{t, i}$ denotes the number of active artificial neurons at timestep $t$, $Y_{t,i}$ represents the number of descendants generated by the $i$-th neuron at that timestep, and $I_t$ denotes the external drive at timestep $t$, assumed to have finite mean and variance.

We then employed the multi-regressive (MR) estimator \cite{wilting2018inferring} to measure the branching parameter $m$, as shown in Fig.~\ref{fig4}(a). Because both the cell state $\boldsymbol{c}_t$ and hidden state $\boldsymbol{h}_t$ were updated at every timestep, the full dynamical system evolved in a state space larger than spanned by $\boldsymbol{h}_t$ alone. However, since only $\boldsymbol{h}_t$ was used in our analysis, this procedure effectively constituted spatial subsampling of the underlying dynamics. The MR estimator was explicitly developed to address such subsampling scenarios and is known to yield unbiased estimates of the branching parameter under these conditions. In addition, thresholding in the avalanche detection procedure suppressed low-amplitude activities, which can likewise be interpreted as an additional form of spatial subsampling. For each input review processed by the LSTM, the hidden state $\boldsymbol{h}_t$ yields an individual activity time series $\{X_t^r\}$, where $r$ indexes the review. The autocorrelation function (ACF) was computed separately for each activity time series and then averaged across all reviews. The MR estimation was subsequently performed by fitting this review-averaged ACF,
\begin{equation}
    \label{equation5}
    C[\tau]=\frac{1}{R}\sum_r \frac{Cov(X_t^r, X_{t+\tau}^r)}{Var(X_t^r)},
\end{equation}
where $R$ denotes the total number of reviews. Hereafter, the term ACF refers to this review-averaged autocorrelation function. The averaged ACF was fitted with an exponential form $bm^{\tau}$ over relatively short time lags up to $\tau=10$, owing to the limited sequence length of $X_t$, where $b$ is a constant. In the left panel of Fig.~\ref{fig4}(b), after optimization, $m$ approached values close to $0.9$ as the hidden-state dimensionality decreased, indicating increasingly critical dynamics in smaller networks. For small networks, $m$ increased monotonically over the course of training and attained its maximum in the vicinity of the optimal epoch. Furthermore, note that the MR estimation method does not impose any requirement for $X_t$ to be represented as integer counts or real-valued quantities. This flexibility allowed $X_t$ to be redefined in terms of the total intensities of the active neurons at each timestep $t$. Specifically, we defined 
\begin{equation}
    \label{equation6}
    X_t = \sum_i |h_{t, i}^{\textrm{norm}}|,
\end{equation}
which eliminates the need for an explicit thresholding procedure. Importantly, the conclusions drawn from this alternative continuous-valued definition of $X_t$ were consistent with those derived from the integer-valued formulation, as evidenced by the agreement between the left and right panels of Fig.~\ref{fig4}(b), reinforcing the robustness of our findings. These observations are qualitatively in agreement with avalanche statistics discussed in the previous section and support the hypothesized driven branching process in Eq.\ref{equation4} as a plausible mechanism underlying LSTM avalanche dynamics.

\section{Long-range dynamics}
\label{Long-range dynamics}
\begin{figure*}[t]
    \centering
    \includegraphics[width=\linewidth]{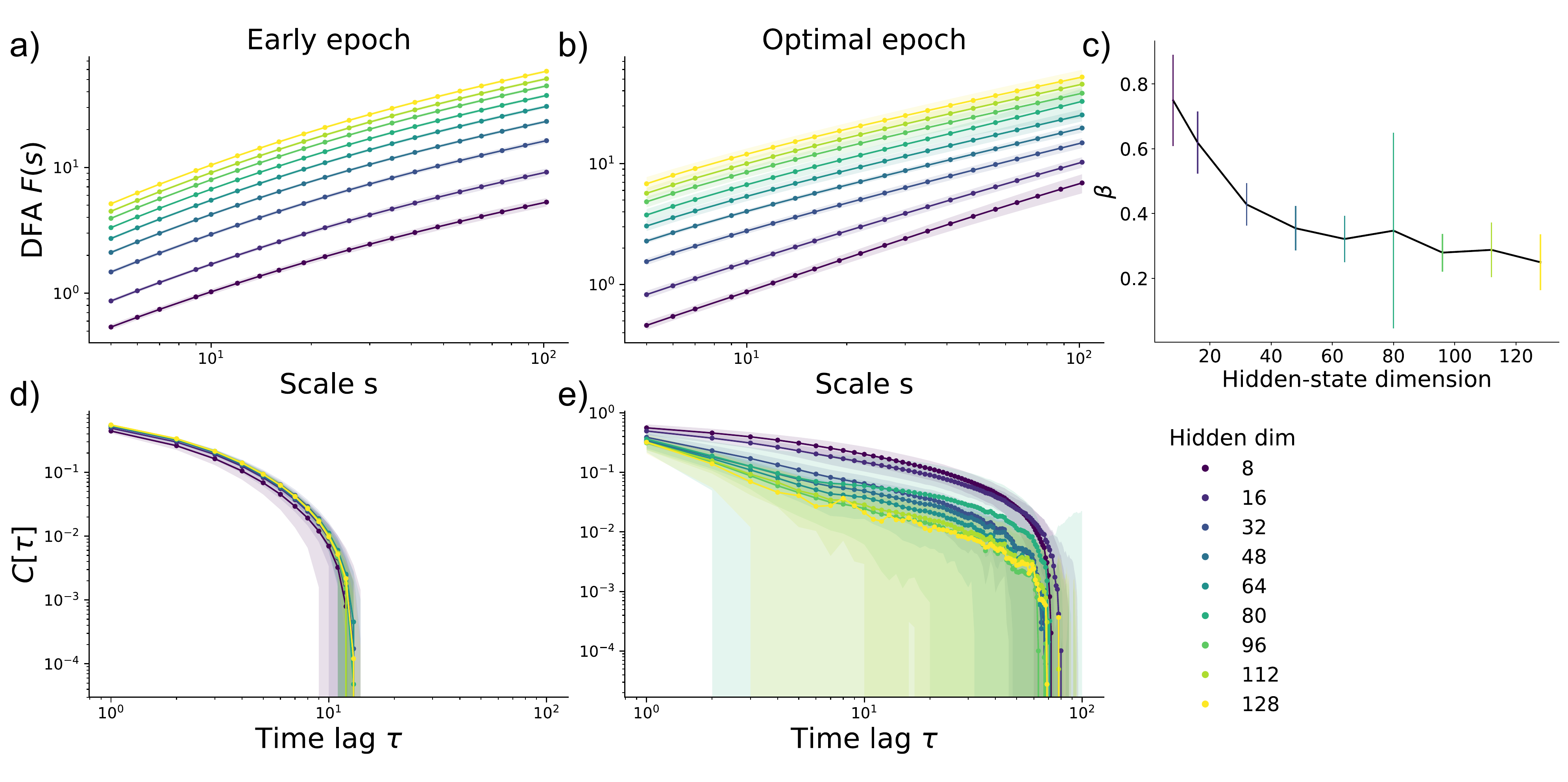}
    \caption{Training-dependent scaling behavior and temporal correlations in LSTM activities. Panels (a) and (b) show DFA fluctuation functions $F(s)$ computed from the discrete-count activity signals $X_t$ as a function of scale $s$ for different hidden-state dimensions, evaluated at an early training epoch (a) and at the optimal epoch (b). Panel (c) presents the spectral exponent $\beta$ of the corresponding $1/f^{\beta}$ noise at the optimal epoch, obtained by fitting $F(s)$ in panel (b), shown as a function of the LSTM hidden-state dimension; error bars are defined as in Fig.~\ref{fig3}. Panels (d) and (e) display the corresponding ACF $C[\tau]$ as a function of time lag $\tau$ at the early (d) and optimal (e) epochs, respectively; shaded regions are defined as in Fig.~\ref{fig4}. At the optimal epoch, the activity exhibits well-defined scale-invariant behavior over multiple window sizes together with an extended correlation length. In contrast, at the early training epoch, $F(s)$ deviates from a single power-law form and the ACF decays rapidly, indicating short-range temporal correlations. Corresponding results for the continuous-valued $X_t$ are provided in Appendix~\ref{Appendix D}.}
    \label{fig5}
\end{figure*}
While the preceding sections established a framework for avalanche analysis in LSTMs, the connection between the inferred branching mechanism and the observed $1/f^{\beta}$ noise remains unsolved. This gap is particularly striking in light of the mean-field sandpile theory, where $1/f^{\beta}$ noise arises from the scale-free duration of avalanches at criticality, while subcritical sandpiles ($m<1$) exhibit a finite characteristic timescale that leads to a Lorentzian power spectral density (PSD)~\cite{per1987self, jensen19891, laurson2005power}. In contrast to these theoretical expectations, prior work \cite{chong2024self} reported robust $1/f^{\beta}$ noise even in larger LSTM networks despite their relatively low branching parameter $m$. This apparent discrepancy suggests that the driven branching model, inferred from the short-range ACF of $X_t$, is insufficient to account for the long-range $1/f^{\beta}$ noise observed in LSTMs.

It is important to note, however, that the activity signal analyzed in Ref.~\cite{chong2024self} is not identical to the definition of $X_t$ adopted here. Specifically, that study considered the summed hidden-state activity $\sum_i h_{t, i}$, whereas in this work $X_t$ was defined either as a discrete count of active neurons or as a continuous-valued aggregation of normalized amplitudes. To ensure a consistent and robust characterization of long-range temporal correlations under this generalized definition, we estimated the spectral exponent $\beta$ of $1/f^{\beta}$ noise in $X_t$ using Detrended Fluctuation Analysis (DFA)~\cite{peng1994mosaic, hardstone2012detrended}. In DFA, for each input review, $X_t^r$ is partitioned into windows of length $s$, locally detrended within each window, and the fluctuation function $F_r(s)$ is defined as the root-mean-square (RMS) magnitude of the detrended residuals, averaged over all windows at scale $s$. The final fluctuation function $F(s)$ is then obtained by averaging $F_r(s)$ across all reviews. For scale-invariant processes, $F(s)$ follows a power-law dependence $F(s)\sim s^{\alpha}$, from which the spectral exponent is obtained via $\beta=2\alpha-1$~\cite{buldyrev1995long}. As shown in Fig.~\ref{fig5}(b), for LSTMs operating near their optimal epoch, the measured $F(s)$ exhibits power-law scaling over approximately two decades in window size $s$. Correspondingly, as shown in Fig.~\ref{fig5}(c), the inferred spectral exponent $\beta$ increases toward values near $0.8$ as the hidden-state dimension decreases, in qualitative agreement with Ref.~\cite{chong2024self}. By contrast, at early training epochs, $F(s)$ is not well described by a single power law and instead displays a crossover with different effective scaling exponents at small and large window sizes, as illustrated in Fig.~\ref{fig5}(a).

To elucidate the origin of the observed $1/f^{\beta}$ noise in $X_t$, it is thus necessary to probe the ACF $C[\tau]$ over substantially longer timescales. In Fig.~\ref{fig5}(e), when evaluated over two decades in time lag $\tau$, $C[\tau]$ computed from $X_t$  at the optimal epoch of all LSTMs exhibits a heavy-tailed behavior that can be well captured by a power law with an exponential cutoff
\begin{equation}
    \label{equation7}
    C[\tau] \sim {\tau}^{-\eta} e^{-\tau/\tau_c},
\end{equation}
where $\eta$ is the power-law exponent, and $\tau_c$ denotes the characteristic cutoff timescale. To mitigate the statistical fluctuations across different realizations, especially at large $\tau$, we fitted this form to the empirical $C[\tau]$, obtained by averaging ACFs across networks trained with different random seeds at the same training epoch and hidden-state dimensionality before fitting the resulting averaged correlation function. This contrasts with the short-range MR analysis, where the branching parameter is fitted separately for each seed and then averaged. The fitted exponent $\eta$ increases systematically with the hidden-state dimensionality, mirroring our earlier observations regarding the dimensional dependence of $\beta$ at the optimal epoch. This correspondence is expected on general grounds, since the scaling form in Eq.~\ref{equation7} directly constrains the high-frequency behavior of the PSD. Specifically, a correlation function of the form $C[\tau] \sim {\tau}^{-\eta} e^{-\tau/\tau_c}$ yields a PSD that is flat for $\omega\ll \omega_c$ and scales as $S(\omega)\sim\omega^{-(1-\eta)}$ for $\omega\gg\omega_c$, with crossover frequency $\omega=1/\tau_c$ (see Appendix \ref{Appendix A} for a derivation). Because $\tau_c$ is typically large in our setting, the experimentally accessible spectrum is dominated by the high-frequency regime, implying $\beta=1-\eta$, in agreement with the empirically observed anti-correlation between $\beta$ and $\eta$. In contrast, $C[\tau]$ at the early epochs displays a pure exponential decay, vanishing over very short time lags. Such an exponential ACF implies a Lorentzian power spectrum that is flat at low frequencies and crosses over to $1/f^2$ at high frequencies, consistent with the two-regime scaling behavior of $F(s)$ observed during early epochs.

\begin{figure}[t]
    \centering
    \includegraphics[width=0.7\linewidth]{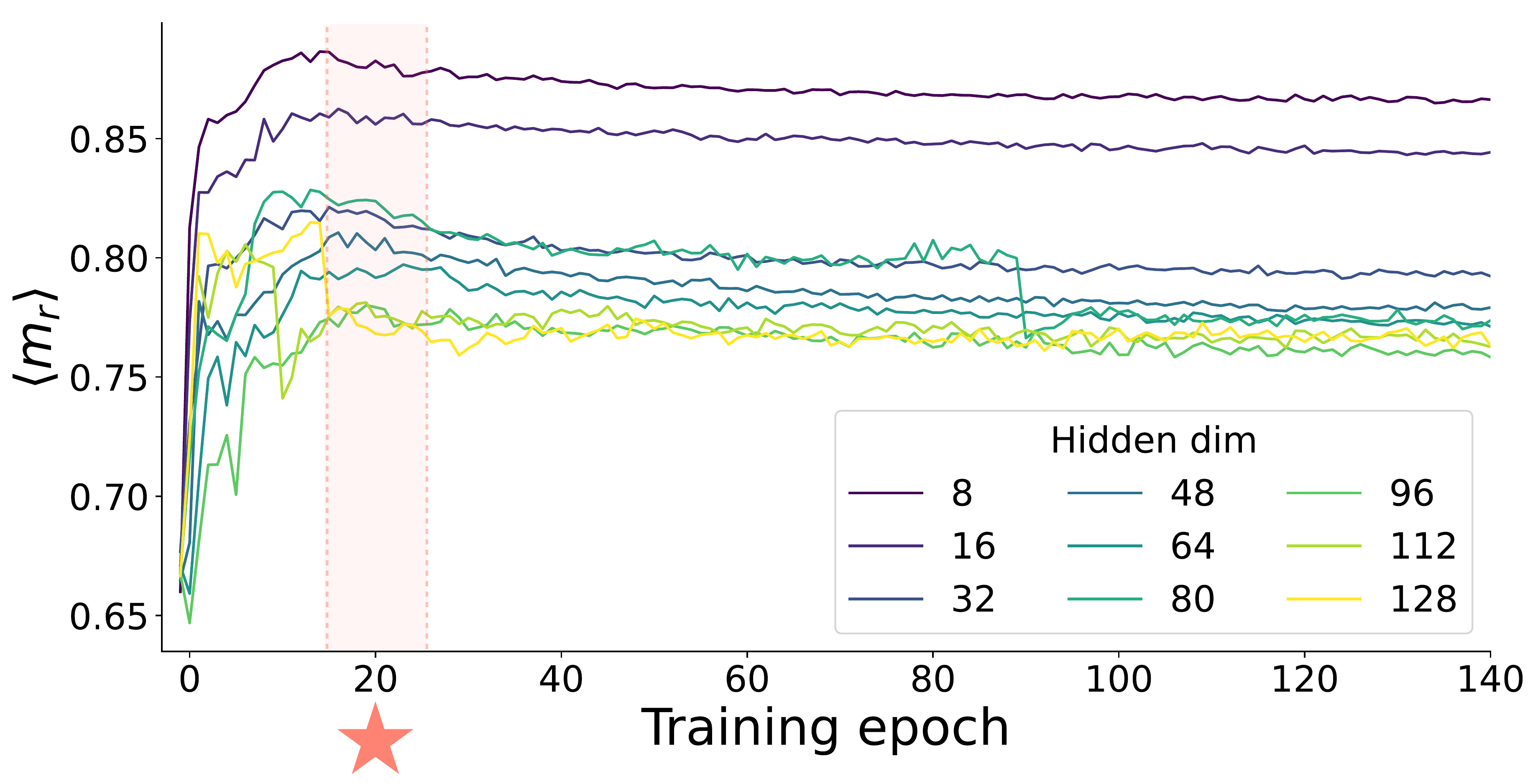}
    \caption{Evolution of the ensemble-averaged branching parameter $\langle m_r \rangle$ as a function of training epoch for different LSTM hidden-state dimensions, inferred from the discrete-count activity signal $X_t$. The red dashed vertical lines and star marker are defined consistently with Fig.~\ref{fig4}. Across all model sizes, $\langle m_r \rangle$ rapidly increases during early training and subsequently stabilizes. Notably, networks with smaller hidden dimensions converge to larger asymptotic values of $\langle m_r \rangle$, exhibiting trends consistent with those shown in Fig.~\ref{fig4}(b).}
    \label{fig6}
\end{figure}

Motivated by the fact that a single branching process cannot generate a genuinely long-ranged temporal correlation, and by the limitation of our preceding analysis in assuming a homogeneous mechanism across different input reviews, we hypothesized a more general mechanism, termed the mixture branching process (MBP), to account for the observed behavior. In the MBP, each review induces its own branching dynamics so that the activity $X_t$ associated with review $r$ evolves with a review-specific branching parameter $m_r$,
\begin{equation}
    \label{equation8}
    X_{t+1}^r = \sum_{i=1}^{X_t^r} Y_{t,i}^r + I_t^r, \quad \langle Y_{t,i}^r \rangle=m_r,
\end{equation}
where $r$ indexes the review. The observed dynamics then arise from the superposition of branching processes with heterogeneous $m_r$, which can effectively reproduce the heavy-tailed decay of $C[\tau]$. Within this framework, the review-specific parameters $m_r$ are sampled from a distribution $W(m_r)$, which can be derived analytically from the scaling form $C[\tau] \sim {\tau}^{-\eta} e^{-\tau/\tau_c}$. The parameters $\eta$ and $\tau_c$ are obtained by the fitting procedure described above. As proven in Appendix \ref{Appendix B}, the analytical form of $W(m_r)$ is
\begin{equation}
    \label{equation9}
    W(m_r) \propto \frac{( -\ln m_r - \lambda_c )^{\eta-1}}{\Gamma(\eta)m_r}\mathbf{1}_{[m_{\min},\,e^{-\lambda_c}]}(m_r),
\end{equation}
where $\Gamma(\cdot)$ denotes the Gamma function, $\lambda_c=1/\tau_c$, and $\mathbf{1}_{[m_{\min},\,e^{-\lambda_c}]}(m_r)$ is a boxcar function that equals unity for $m_r \in [m_{\min},\,e^{-\lambda_c}]$ and vanishes otherwise. The introduction of a lower bound $m_{min}$ is required to prevent an unphysical divergence associated with extremely small $m_r$. Importantly, the specific choice of $m_{min}$ affects only short-timescale behavior and does not alter the heavy-tailed regime of $C[\tau]$, which is the primary focus of our long-range analysis. In the calculations that follow, we therefore fix $m_{min}=0.6$. The upper bound $m_c=e^{-\lambda_c}$ arises directly from the exponential cutoff in ACF. Motivated by this distribution $W(m_r)$, we further defined an ensemble-averaged branching parameter,
\begin{equation}
    \label{equation10}
    \langle m_r \rangle = \int_0^1m_rW(m_r)dm_r,
\end{equation}
which captures the effective contribution of heterogeneous branching dynamics to the observed long-range correlations. This quantity admits a closed-form expression,
\begin{equation}
    \label{equation11}
    \langle m_r \rangle = \eta m_c \frac{\gamma(\eta, \ln\frac{m_c}{m_{min}})}{(\ln\frac{m_c}{m_{min}})^{\eta}},
\end{equation}
where $\gamma(\cdot)$ denotes the lower incomplete Gamma function. To characterize networks across different training epochs and hidden-state dimensionalities, the fitted parameters $\eta$ and $\tau_c$ were substituted into Eq.~\ref{equation11} to obtain the corresponding value of $\langle m_r \rangle$. In Fig.~\ref{fig6}, the resulting ensemble-averaged branching parameter exhibits trends that closely parallel those of the branching parameter $m$ inferred from the short-range ACFs across networks and training epochs, indicating a consistent characterization of LSTM dynamics across both short and long timescales. 

Within the MBP framework, the avalanche size distribution $P(S)$ can be derived analytically under certain assumptions. Specifically, assuming that the external drive $I_t^r$ is moderate and its effect can be negligible due to the thresholding procedure, each review-specific branching process with a fixed $m_r<1$ generates avalanches that follow a distribution of the form 
\begin{equation}
    \label{equation12}
    P(S, m_r) \sim S^{-3/2}exp[-\frac{S}{S_c(m_r)}] \qquad S \rightarrow \infty,
\end{equation}
where $S_c$ denotes the characteristic scale of the exponential portion of the distribution~\cite{christensen1993sandpile}. The scale-free behavior of $P(S)$ reported in Sec.~\ref{Avalanches} arises only after averaging over the heterogeneous distribution $W(m_r)$. As shown in Appendix \ref{Appendix C}, the superposition of branching processes whose parameters are distributed according to Eq. \eqref{equation9} yields an avalanche size distribution of the form
\begin{equation}
    \label{equation13}
    P(S) \sim S^{(-\frac{3}{2} + \eta)} exp{[-\frac{S(1-m_c)^2}{C}]},
\end{equation}
where $C$ is a constant, and the power-law exponent $\tau_s = \frac{3}{2} + \eta$ is directly controlled by the correlation exponent $\eta$. During the optimal epochs of the smallest LSTM network, the upper-bound cutoff $m_c$ in the weight distribution $W(m_r)$ approaches unity, leading to an extended power-law regime, consistent with the avalanche statistics observed empirically near these epochs. This explains why the measured avalanche exponent $\tau_s$ is usually found to be larger than $\frac{3}{2}$ in small networks. Moreover, this result reinforces the unifying role of the MBP framework: both the long-range temporal correlations and the avalanche size distributions originate from the same underlying heterogeneity in branching dynamics, providing a coherent explanation for critical-like behavior across multiple statistical observables.

\section{Discussion}
\label{Discussion}
Our study demonstrates that artificial recurrent neural networks like the LSTM networks exhibit criticality-like dynamics when it is well trained, analogous to neuronal avalanches observed in biological neural systems. Through a combination of avalanche statistics, branching-process analysis, and spectral characterization, we show that multiple hallmarks of critical dynamics consistently emerge in smaller models near their optimal training epochs. The simultaneous appearance of these signatures across independent analytical frameworks provides converging evidence that LSTMs can transiently organize into a regime reminiscent of criticality during learning. 

Building on this observation, a central insight of this work is that the observed critical-like behavior reflects a unified dynamical picture spanning both short-range and long-range temporal organization. At short timescales, the dynamics are governed by an effective branching process characterized by a branching parameter close to unity, which controls the propagation and termination of local activity cascades and gives rise to avalanche statistics. At longer timescales, heterogeneity in the effective branching parameters across different input conditions gives rise to a mixture of branching processes, whose superposition naturally generates long-range temporal correlations and $1/f^{\beta}$ spectral behavior. In this sense, short-range avalanche dynamics and long-range memory effects are not independent phenomena but arise from two complementary branching mechanisms operating at different temporal scales within the same system.

Furthermore, our results indicate that such critical behavior is not an intrinsic consequence of the LSTM architecture itself. Instead, it emerges from an interplay between training dynamics and model capacity. Larger models, despite their greater expressive power, tend to operate further from this critical regime, exhibiting lower effective branching parameters and weaker long-range temporal correlations. This suggests that overparameterization may suppress the fine balance between amplification and dissipation required for critical-like dynamics, which could be related to the delayed optimization observed in larger networks (Fig.~\ref{fig11}), whereas smaller models are naturally driven toward this balance during training. In this sense, criticality appears as an emergent, capacity-dependent property rather than a built-in architectural feature. The training epoch plays a similarly crucial role. Critical signatures are most pronounced near the optimal epoch, where generalization performance is maximized, and diminish at early epochs when dynamics are strongly subcritical and dominated by short timescales. This temporal evolution highlights a close connection between learning, dynamical organization, and criticality. Rather than being static, the dynamical state of an LSTM evolves throughout training, transiently approaching criticality as representations become maximally informative and flexible. 

Beyond LSTMs, our findings suggest that branching parameters may provide a useful, architecture-agnostic descriptor of sequential neural network dynamics, such as Transformers~\cite{vaswani2017attention}, MAMBA~\cite{gu2024mamba} and DeltaNet~\cite{schlag2021linear}. Because the branching parameter captures how activity propagates over time, it offers a compact and interpretable measure of dynamical regime that is largely independent of specific architectural details. As such, it may serve as a unifying indicator for comparing different recurrent architectures, training protocols, or capacity regimes. More broadly, the emergence of critical-like dynamics in artificial networks reinforces the idea that criticality may be a general organizing principle for efficient information processing, arising naturally in systems that balance stability and adaptability through learning.

\section{Data availability}
The data that support the findings of this article are openly available~\cite{ren_dataset_2026}. The code used to reproduce the results is available in a GitHub repository~\cite{ren_code_2026}.

\appendix
\setcounter{figure}{0}
\renewcommand{\thefigure}{A\arabic{figure}}

\section{Relationship between $\beta$ and $\eta$}
\label{Appendix A}
In this appendix, we derive the relationship between the spectral exponent $\beta$ and the correlation exponent $\eta$ for $X_t$ whose ACF exhibits power-law decay with an exponential cutoff
\begin{equation}
    \label{A1} 
    C[\tau] \sim \tau^{-\eta} e^{-\tau/\tau_c}.
\end{equation}
where $0<\eta<1$ characterizes long-range temporal correlations and $\tau_c$ denotes a finite correlation timescale. For a stationary process, the PSD is related to the ACF through the Wiener-Khinchin theorem. Assuming that $X_t$ is stationary, its PSD is given by
\begin{equation}
    \label{A2}
    \begin{aligned}
        S(\omega) 
        &= \sum_{\tau=-\infty}^{\infty} R[\tau] e^{-j \omega \tau} \, d\tau \\
        & \approx \sum_{\tau=1}^{\infty} \tau^{-\eta} e^{-\tau/\tau_c} \cos(\tau\omega) \\
        &\approx \tau_c^{1-\eta}\int_{0}^{\infty} x^{-\eta} e^{-x} \cos(\omega \tau_c x)\, dx,
    \end{aligned}
\end{equation}
where $R[\tau] = \mathbb{E}[X_tX_{t+\tau}]$, which differs from the ACF $C[\tau]$ defined earlier, but does not affect the scaling behavior at finite frequencies. In the second line, we exploit the even symmetry of the autocorrelation function, and in the third line we approximate the discrete sum by a continuum integral, which is justified for sufficiently large $\tau_c$. The integral in Eq.~\ref{A2} can be rewritten in complex form as
\begin{equation}
    \label{A3}
    \begin{aligned}
        \tau_c^{1-\eta}\int_{0}^{\infty} &x^{-\eta} e^{-x} \cos(\omega \tau_c x)dx \\
        &= \tau_c^{1-\eta}\Re \left[\int_{0}^{\infty} x^{-\eta}e^{-(1 - j\omega \tau_c)x}dx\right],
    \end{aligned}
\end{equation}
For $0<\eta<1$, the integral converges and can be evaluated using the identity
\begin{equation}
    \label{A4}
    \int_{0}^{\infty} x^{-\eta} e^{-ax}\, dx= \Gamma(1-\eta)a^{\eta-1},  \qquad \Re(a) > 0.
\end{equation}
Applying this result with $a = 1 - j\omega \tau_c$ yields
\begin{equation}
    \label{A5}
    S(\omega)\sim
    \tau_c^{1-\eta}
    \Re\left[
    \frac{\Gamma(1-\eta)}{(1 - j\omega \tau_c)^{1-\eta}}
    \right].
\end{equation}
For small frequencies where $\omega k_c \ll 1$, the PSD tends to a constant:
\begin{equation}
    \label{A6}
    S(\omega) \sim (\tau_c)^{1-\eta}\Gamma(1-\eta)
    \to \text{constant}.
\end{equation}
In the large-frequency regime $\omega \tau_c \gg 1$, we approximate
\begin{equation}
    \label{A7}
    \begin{aligned}
        (1 - j\omega \tau_c)^{-(1-\eta)}
        &\approx (\omega \tau_c e^{-j\frac{\pi}{2}})^{-(1-\eta)} \\
        &= (\omega \tau_c)^{-(1-\eta)} e^{ i(1-\eta)\pi/2}.
    \end{aligned}
\end{equation}
Thus, the PSD becomes
\begin{equation}
    \label{A8}
    \begin{aligned}
        S(\omega) 
        &\sim \tau_c^{1-\eta} (\omega \tau_c)^{\eta-1}\Gamma(1-\eta)\sin\!\left( \frac{\pi\eta}{2} \right) \\
        &\sim\Gamma(1-\eta)\sin\!\left(\frac{\pi\eta}{2}\right)\,\omega^{\eta-1}.
    \end{aligned}
\end{equation}
Defining the spectral exponent $\beta$ via $S(\omega)\sim \omega^{-\beta}$, we obtain the relation
\begin{equation}
    \label{A9}
    \beta + \eta=1.
\end{equation}



\section{Derivations of long-range dynamics}
\label{Appendix B}
In this section, we derive the weight distribution $W(m_r)$ for the MBP whose ACF follows a power-law decay with an exponential cutoff. We then obtain the analytical expression for the ensemble-averaged branching parameter $\langle m_r \rangle$ from the derived distribution $W(m_r)$, and illustrate why it is necessary to introduce a lower-bound truncation $m_{min}$ for $W(m_r)$. Finally, we quantify the error introduced by this truncation. 

\subsection{Derivation of $W(m_r)$}
Since the MBP is mathematically described by Eq.~\ref{equation8}, the ACF $C[\tau]$, averaged over all $X_t$ corresponding to different test reviews, can be written as
\begin{equation}
    \label{B1}
    C[\tau] = \int_0^1 W(m_r)m_r^\tau\, dm_r.
\end{equation}
We introduce the variable transformation $m_r = e^{-\lambda}$, , which maps the interval $m_r \in (0, 1)$ to $\lambda \in (0, \infty)$. Under this change of variables, Eq.~\ref{B2} becomes
\begin{equation}
    \label{B2}
    \begin{aligned}
        C[\tau] &= \int_0^\infty W(e^{-\lambda})e^{-\tau\lambda}e^{-\lambda}\,d\lambda \\
        &= \int_0^\infty \tilde{W}(\lambda)e^{-\tau\lambda}\, d\lambda,
    \end{aligned}
\end{equation}
where we have defined $\tilde{W}(\lambda)=W(e^{-\lambda})e^{-\lambda}$. The above expression shows that $C[\tau]$ is the Laplace transform of $\tilde{W}(\lambda)$. A standard Laplace transform pair for $\Re(\eta)>0$ is given by
\begin{equation}
    \label{B3}
    \int_{0}^{\infty} \frac{\lambda^{\eta-1}}{\Gamma(\eta)}e^{-\tau\lambda}\, d\lambda= \tau^{-\eta}.
\end{equation} 
Furthermore, a shift in $\lambda$–space by $\lambda_c= \frac{1}{\tau_c}$ introduces an exponential factor in the transform,
\begin{equation}
    \label{B4}
    \int_{\lambda_c}^{\infty}\frac{(\lambda-\lambda_c)^{\eta-1}}{\Gamma(\eta)}e^{-\tau\lambda}\, d\lambda=e^{-\tau\lambda_c}\tau^{-\eta}.
\end{equation}        
Comparing this expression with the assumed form of the ACF $C[\tau] \sim \tau^{-\eta}e^{-\tau\lambda_c}$, we obtain
\begin{equation}
    \label{B5}
    \tilde{W}(\lambda) \propto \frac{(\lambda - \lambda_c)^{\eta-1}}{\Gamma(\eta)} H(\lambda - \lambda_c),
\end{equation}
where $H(\cdot)$ denotes the Heaviside step function. Transforming back to the original variable $m_r$, the weight distribution is given by
\begin{equation}
    \label{B6}
    W(m_r) = \frac{1}{Z}\frac{( -\ln m_r - \lambda_c )^{\eta-1}}{\Gamma(\eta)m_r}H(-\ln m_r - \lambda_c).
\end{equation}
where $Z$ ensures proper normalization of $W(m_r)$ over the domain of $m_r \in[0, 1]$. From the analytical form, the lower bound in $m_r$-space follows directly as 
\begin{equation}
    \label{B7}
    m_c = e^{-\frac{1}{\tau_c}}.
\end{equation}
Consequently, $W(m_r)$ is non-zero only for $0 < m_r < m_c$.

\begin{figure*}
    \centering
    \includegraphics[width=0.9\linewidth]{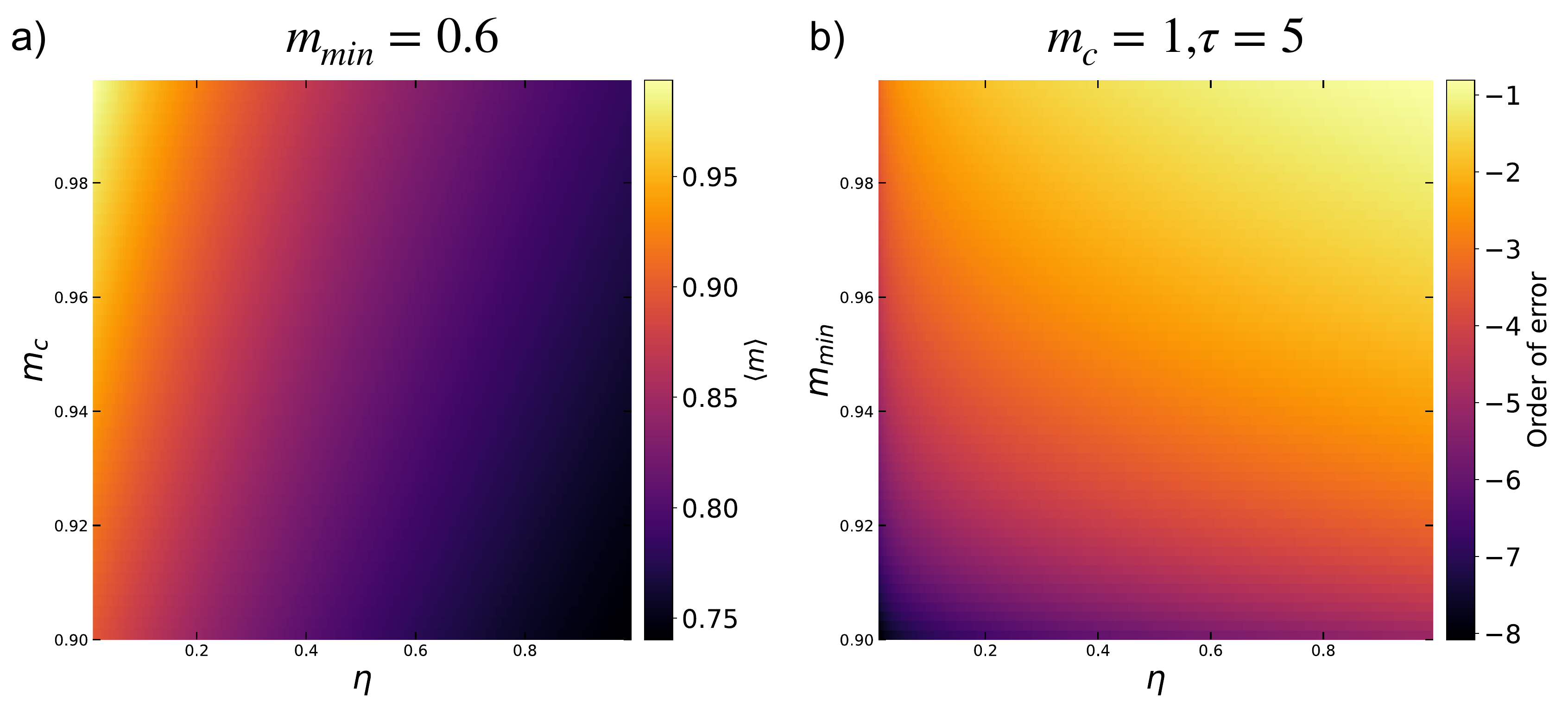}
    \caption{(a) Ensemble-averaged branching parameter $\langle m_r \rangle$ as a function of the correlation exponent $\eta$ and the upper bound $m_c$ for a fixed lower bound $m_{min}=0.6$. (b) Order of magnitude of the relative truncation error $Q(\eta, \tau d)$, shown as a function of $\eta$ and $m_{min}$ for $m_{min}=1$ and $\tau=5$. The error is of order $10^{-k}$ across the explored parameter range, demonstrating that the truncation-induced correction decays rapidly for large $\tau d$ and does not affect the long-range temporal behavior of the ACF $C[\tau]$.}
    \label{fig7}
\end{figure*}

\subsection{The ensemble-averaged branching parameter}
Given the weight distribution $W(m_r)$, a natural quantity to consider is an ensemble-averaged branching parameter that captures the heterogeneous branching dynamics across reviews,
\begin{equation}
    \label{B8}
    \langle m_r \rangle
    = \displaystyle \int_0^1 m_r\, W(m_r)\, dm_r.
\end{equation}
However, the normalization integral of $Z$ diverges. This divergence follows from the asymptotic behavior of the weight distribution,
\begin{equation}
    \label{B9}
    W(m_r)\sim \frac{\ln(1/m_r)^{\eta-1}}{\Gamma(\eta)m_r} \qquad m_r \rightarrow 0^{+},
\end{equation}
which produces an unphysical divergence as $m_r$ approaches zero. This pathology reflects the excessive weight assigned to extremely small branching parameters, which do not correspond to meaningful dynamical realizations. Importantly, the ACF $C[\tau]$ retains its heavy-tailed form even if a lower-bound cutoff $m_{min}$ is introduced and only the remaining portion of $W(m_r)$ is retained. Motivated by this observation, we define a truncated weight distribution,
\begin{equation}
    \label{B10}
    W(m_r) = \frac{1}{Z}\frac{(-\ln m_r- \lambda_c)^{\eta-1}}{\Gamma(\eta)m_r}\mathbf{1}_{[m_{min}, m_c]}(m_r).
\end{equation}
This truncation removes the spurious divergence at $m_r \rightarrow 0$ while preserving the long-range temporal correlations encoded in the ACF. Under this truncation, the normalization constant $Z$ becomes
\begin{equation}
    \begin{aligned}
        Z
        &= \int_{\lambda_c}^{\lambda_{max}}\frac{(\lambda - \lambda_c)^{\eta-1}}{\Gamma(\eta)}\, d\lambda \\
        &= \frac{(\lambda_{max} - \lambda_c)^{\eta}}{\eta\Gamma(\eta)} \\
        &= \frac{d^{\eta}}{\eta\Gamma(\eta)},
    \end{aligned}
\end{equation}
where $\lambda_{max}=-\ln m_{min}$, and we have defined $d = \lambda_{max} - \lambda_c$. Substituting the truncated distribution into Eq.~\ref{B8}, the ensemble-averaged branching parameter $\langle m_r \rangle$ is given by
\begin{equation}
    \begin{aligned}
        \int_0^1 m_r W(m_r) dm_r
        &= \frac{1}{\Gamma(\eta)Z} \int_{\lambda_c}^{\lambda_{max}} (\lambda - \lambda_c)^{\eta-1}e^{-\lambda}d\lambda \\
        &= \frac{e^{-\lambda_c}}{\Gamma(\eta)Z} \int_{0}^{d} u^{\beta-1}e^{-u}du \\
        &= \frac{e^{-\lambda_c}}{\Gamma(\eta)Z} \gamma(\eta, d),
    \end{aligned}
\end{equation}
where $\gamma(\eta, d)$ denotes a lower incomplete Gamma function. After straightforward simplification, the ensemble-averaged branching parameter is obtained as
\begin{equation}
    \begin{aligned}
        \langle m_r \rangle
        & = \frac{\eta e^{-\lambda_c}}{d^{\eta}}\gamma(\eta, d) \\
        &= \eta m_c \frac{\gamma(\eta, \ln\frac{m_c}{m_{min}})}{(\ln\frac{m_c}{m_{min}})^{\eta}}.
    \end{aligned}
\end{equation}
A visualization of $\langle m_r \rangle$ as a function of $\eta$ and $m_c$ for $m_{min}=0.6$ is shown in Fig.~\ref{fig7}(a). Within a reasonable range of $m_{min}$ the qualitative dependence of $\langle m_r \rangle$ on these parameters remains unchanged.

\subsection{Error of truncation}
The error introduced by the truncation can be quantified by comparing the ACF computed from the untruncated weight distribution $W(m_r)$ with that obtained from the truncated distribution. The ACF prior to truncation is given by Eq.~\ref{A1}. The truncated ACF is obtained by substituting Eq.~\ref{B10} into Eq.~\ref{B1}, yielding
\begin{equation}
    \begin{aligned}
        C_{trunc}[\tau]
        &\propto \frac{1}{\Gamma(\eta)} \int_{\lambda_c}^{\lambda_{max}} (\lambda - \lambda_c)^{\eta-1}e^{-\tau\lambda}d\lambda \\
        &= \frac{e^{-\tau\lambda_c}\tau^{-\eta}}{\Gamma(\eta)} \int_{0}^{\tau d} u^{\eta-1}e^{-u}du \\
        &= e^{-\tau\lambda_c}k^{-\eta} \frac{\gamma(\eta, \tau d)}{\Gamma(\eta)} \\
        &= e^{-\tau\lambda_c}k^{-\eta} P(\eta,\tau d),
    \end{aligned}
\end{equation}
where $\gamma(\eta, \tau d)$ is the lower incomplete Gamma function, and $P(\eta, \tau d)$ denotes the regularized incomplete Gamma function. The relative error induced by the truncation is therefore given by
\begin{equation}
    err = \frac{C[\tau] - C_{trunc}[\tau]}{C[\tau]} = Q(\eta, \tau d),
\end{equation}
where $Q(\eta, \tau d) = 1 - P(\eta, \tau d)$ denotes the regularized upper incomplete Gamma function. This expression shows that the truncation error decays rapidly for $\tau d \gg 1$, confirming that the long-range behavior of the ACF is preserved. This behavior is explicitly demonstrated in Fig.~\ref{fig7}(b), which shows that $Q(\eta, \tau d)$ remains small for $\tau=5$.



\section{Avalanche size distribution in MBP}
\label{Appendix C}
In this appendix, we derive the asymptotic form of the avalanche size distribution $P(S)$ induced by the weight distribution $W(m_r)$ in the MBP theory. 
For each single subcritical branching process with the review-specific branching parameter $m_r$, the avalanche size distribution is given by
\begin{equation}
    \begin{aligned}
        P(S, m_r) \sim S^{-3/2}exp[-\frac{S}{S_c(m_r)}] &\qquad S\rightarrow\infty,  \\
        S_c(m_r) \sim \frac{1}{(1-m_r)^2} &\qquad m_r \rightarrow 1,
    \end{aligned}
\end{equation}
Given the weight distribution $W(m_r)$, the total avalanche size distribution of the MBP should be
\begin{equation}
    P(S) = \int_0^1 W(m_r)P(S, m_r)dm_r,
\end{equation}
where we  can apply Eq.\ref{B10} for $W(m_r)$. We introduce the shifted variables
\begin{equation}
    \delta = 1 - m_c \qquad \epsilon = m_c - m.
\end{equation}
The avalanche-size distribution can be expressed with new variables
\begin{equation}
    \begin{aligned}
        &P(S) 
        \sim S^{-3/2}\int_{0}^1 W(m)\exp\left[-\frac{S(1-m)^2}{C}\right] dm \\
        &= S^{-3/2}\int_{0}^{\epsilon_{max}} W(m_c - \epsilon)\exp\left[-\frac{S(\delta+\epsilon)^2}{C}\right]d\epsilon,
    \end{aligned}
\end{equation}
where $\epsilon_{max} = m_c - m_{min}$ denotes the upper bound of the integration since $W(m_r)$ is only valid in the interval $m \in [m_{min}, m_c]$, and $C$ is a constant. Using $\lambda_c = -\ln m_c$,
\begin{equation}
    -\ln m - \lambda_c= \frac{\epsilon}{m_c} + O(\epsilon^2) \qquad \epsilon\rightarrow0^{+}.
\end{equation}
Plug the equation into the weight distribution yields
\begin{equation}
    W(m_c - \epsilon) \sim \frac{\epsilon^{\eta-1}}{\Gamma(\eta)m_c^{\eta}} \qquad \epsilon\rightarrow0^{+}.
\end{equation}
In the large-$S$ limit, the upper bound of the integral $\epsilon_{max}$ can be extended to infinity, yielding
\begin{equation}
    P(S) \sim S^{-3/2}\int_{0}^{\infty}\epsilon^{\eta-1}\exp\left[-\frac{S(\delta+\epsilon)^2}{C}\right] d\epsilon.
\end{equation}
Factoring out the dominant exponential term and neglecting the quadratic term in the exponent gives
\begin{equation}
    \begin{aligned}
        P(S)
        &\sim S^{-3/2}\exp\left(-\frac{S\delta^2}{C}\right)\int_{0}^{\infty} \epsilon^{\eta-1} e^{\frac{-2S\delta\epsilon}{C}} d\epsilon \\
        &=S^{-3/2}\exp\left(-\frac{S\delta^2}{C}\right)\Gamma(\eta)\left(\frac{2S\delta}{C}\right)^{-\eta}.
    \end{aligned}
\end{equation}
Combining all factors, we obtain the asymptotic avalanche-size distribution
\begin{equation}
    P(S) \sim S^{-(\frac{3}{2}+\eta)}\exp\left[-\frac{S(1-m_c)^2}{C}\right],
\end{equation}
where $m_c = e^{-1/\tau_c}$. This result shows that the MBP produces a power-law avalanche-size distribution with exponent $\frac{3}{2}+\eta$, followed by an exponential cutoff controlled by the distance $1-m_c$ from criticality.

\section{Additional details}
\label{Appendix D}
\subsection{Excluded Network Realizations}
\begin{figure}
    \centering
    \includegraphics[width=0.9\linewidth]{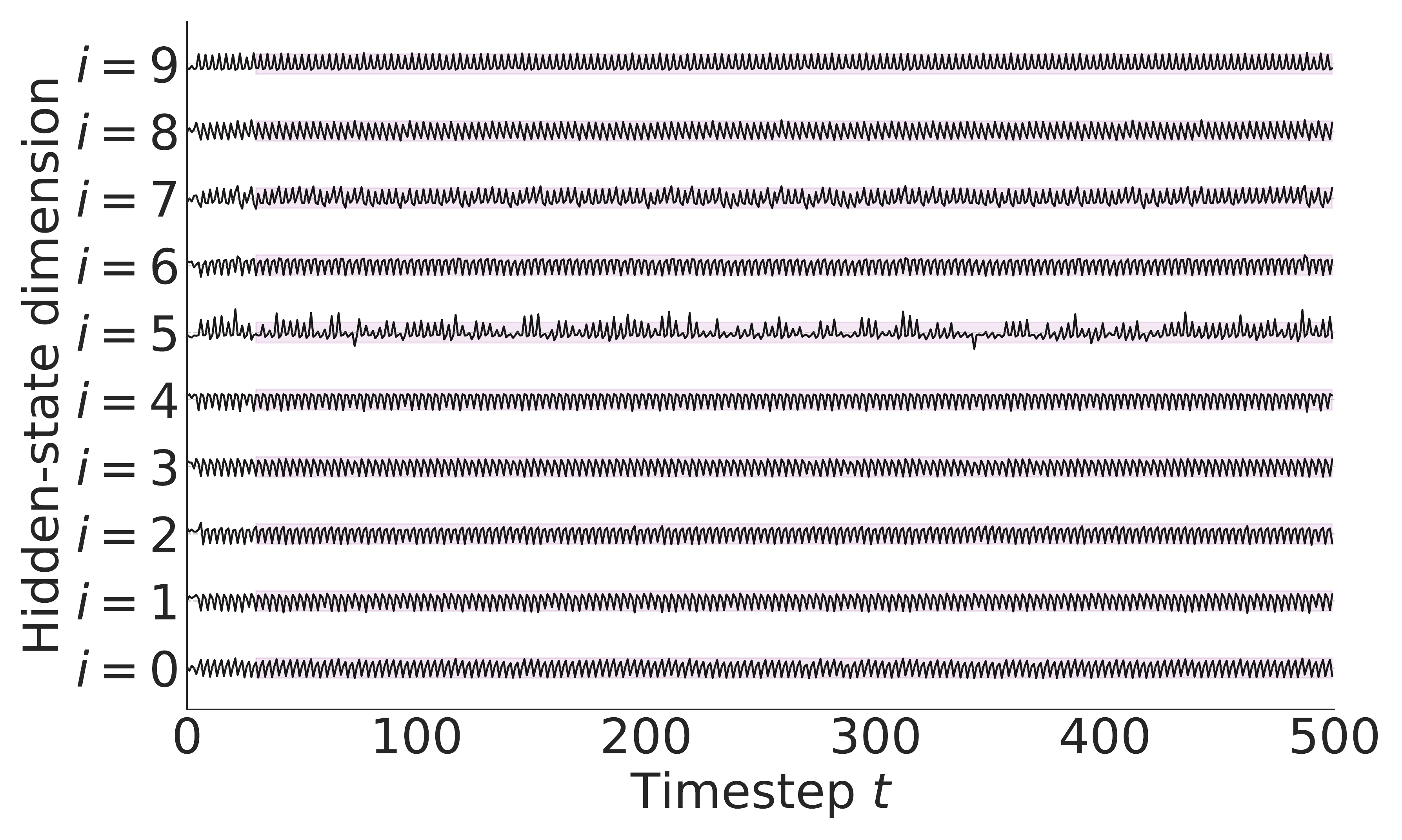}
    \caption{Time series of selected normalized hidden-state $\boldsymbol{h}_t$ from a trained LSTM with hidden-state dimensionality $96$, evaluated on a single test review. For clarity, only ten out of the $96$ hidden-state dimensions are shown. The displayed components exhibit pronounced, synchronized fast oscillations.}
    \label{fig8}
\end{figure}

A small fraction of trained networks was excluded from the avalanche and branching analysis due to pathological behaviors observed during inference. The first exclusion criterion was numerical instability, manifested as NaN values in the hidden states $\boldsymbol{h}_t$. The second criterion was the emergence of persistent fast oscillations in the hidden states $\boldsymbol{h}_t$, synchronized across all hidden-state dimensions, as illustrated in Fig.~\ref{fig8}. In this case, the hidden-state vector repeatedly alternates between two distinct points in the state space across successive timesteps, resulting in a pronounced periodic pattern. Such behavior is indicative of a highly ordered dynamical regime and is typically observed at early training epochs. While this oscillatory behavior may be related to bifurcation-like phenomena in recurrent dynamical systems~\cite{pascanu2013difficulty}, a detailed investigation of its origin is beyond the scope of the present work. As a consequence, the corresponding autocorrelation function $C[\tau]$ displays strong periodic structure rather than the exponential decay assumed in the branching-process framework, rendering these cases unsuitable for our analysis. Both types of pathological behavior occur more frequently in networks with larger hidden-state dimensionality. Nevertheless, for each hidden-state dimension, no more than four out of 28 trained networks were affected, and excluding these cases does not affect the statistical robustness or validity of our conclusions.

\begin{figure*}
    \centering
    \includegraphics[width=\linewidth]{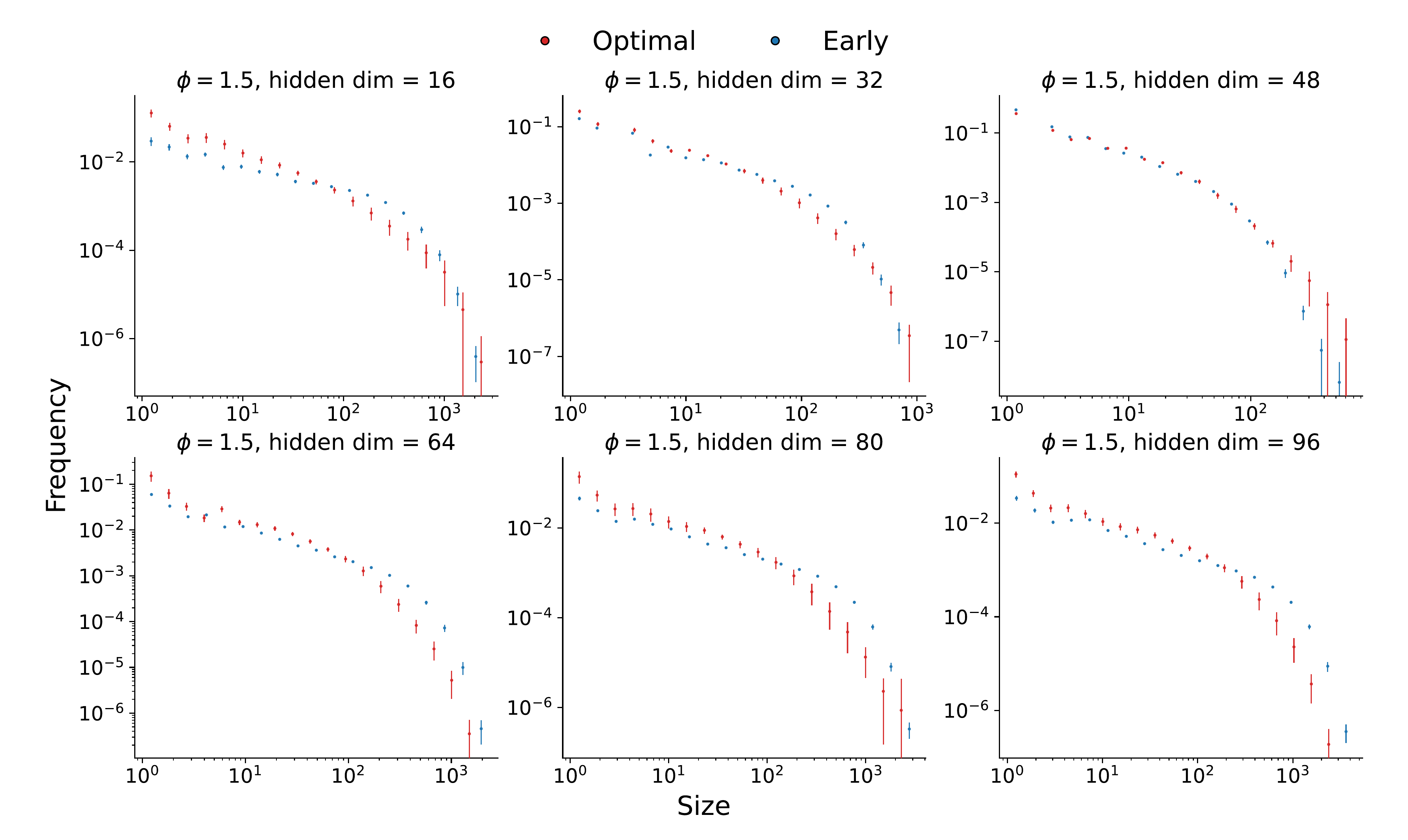}
    \caption{Avalanche size distributions $P(S)$ for LSTM networks with hidden-state dimensionalities $16–96$ at $\phi=1.5$, shown for early (blue) and optimal (red) training epochs. Error bars indicate variability across random seeds.}
    \label{fig9}
\end{figure*}
\subsection{Avalanche statistics in larger LSTM networks}
In the main text, the avalanche size distribution $P(S)$ for a hidden-state dimensionality of $128$ is shown as a representative example, as illustrated in Fig.~\ref{fig3}(d). Here, we present corresponding results for additional larger networks to demonstrate that the observed subcritical behavior is not specific to that particular network size. As shown in Fig.~\ref{fig9}, LSTM networks with hidden-state dimensionalities ranging from 16 to 96 exhibit avalanche size distributions that deviate from power-law scaling at both early and optimal training epochs, indicating persistent subcritical dynamics. These results confirm that the subcritical avalanche behavior observed in large networks is robust across different hidden-state dimensionalities.

\subsection{Analysis Using Continuous-Valued Activity Signals}
\begin{figure*}
    \centering
    \includegraphics[width=\linewidth]{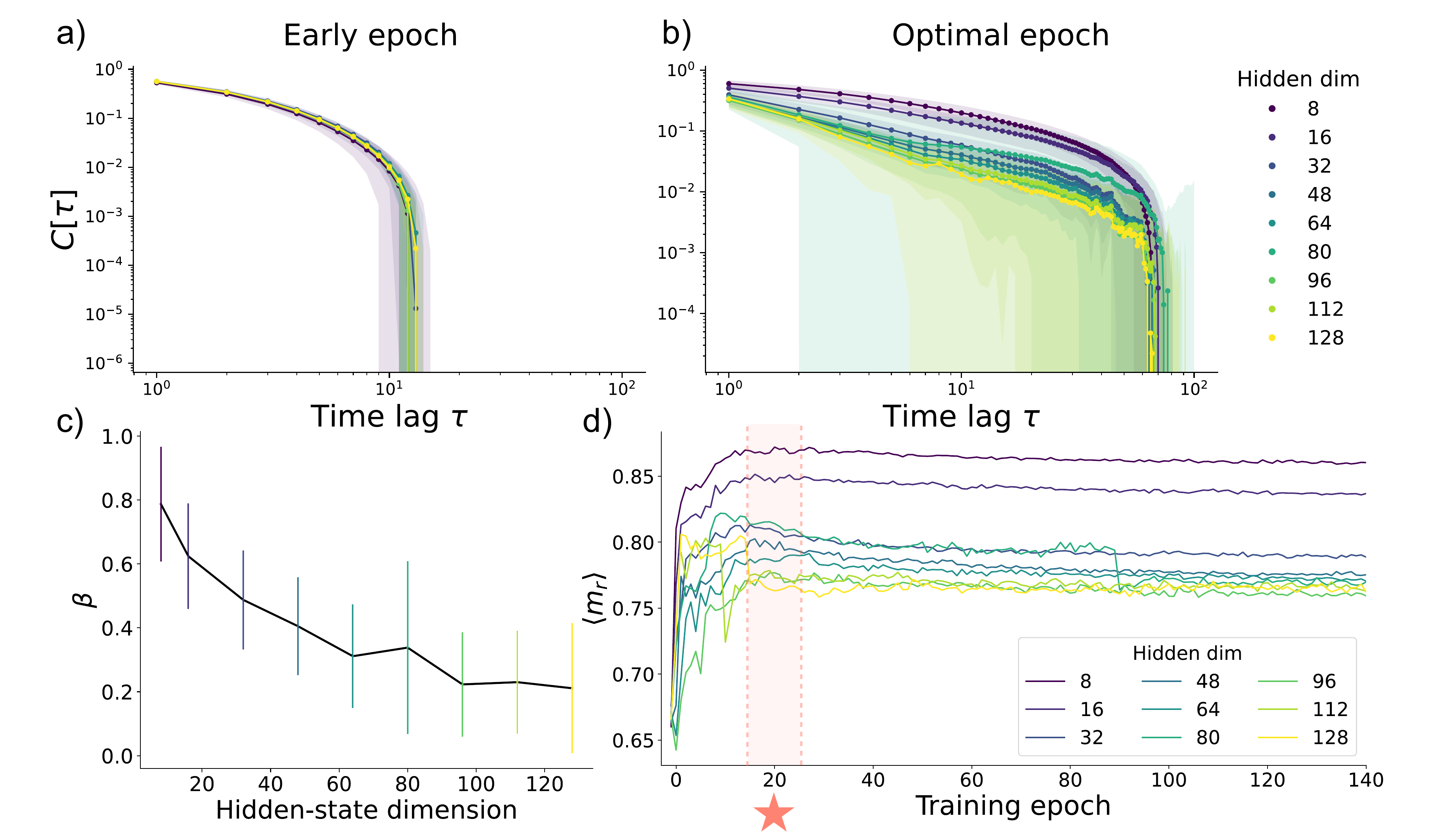}
    \caption{Temporal correlations and branching statistics for continuous-valued activity signals. (a,b) ACF $C[\tau]$ of the continuous-valued activity $X_t$ at the early (a) and optimal (b) training epochs for different hidden-state dimensionalities. (c) Spectral exponent $\beta$ of the corresponding $1/f^{\beta}$ noise at the optimal epoch, shown as a function of hidden-state dimensionality. (d) Evolution of the ensemble-averaged branching parameter $\langle m_r \rangle$ over training epochs for different hidden-state dimensionalities. Results obtained using discrete-count activity signals are shown in the main text and display qualitatively similar behavior.}
    \label{fig10}
\end{figure*}
To assess the robustness of our results with respect to the definition of the activity signal, we repeated the temporal-correlation and branching analyses using a continuous-valued definition of the activity $X_t$, which removes the need for explicit thresholding and binarization of the hidden states. Fig.~\ref{fig10} summarizes the results obtained from this continuous-valued activity signal. The ACFs, spectral exponents $\beta$, and ensemble-averaged branching parameters $\langle m_r \rangle$ exhibit trends that closely mirror those observed for the discrete-count activity signals presented in the main text. In particular, small networks operate closer to criticality at the optimal epoch, while larger networks remain subcritical, and training drives the system toward longer temporal correlations. These results demonstrate that our conclusions are not sensitive to the specific choice of activity definition and are therefore robust.

\bibliographystyle{unsrtnat}
\bibliography{references}  






\end{document}